\documentstyle[12pt]{article}
 \newcommand{\be}{\begin{equation}}
\newcommand{\ee}{\end{equation}}
\newcommand{\bea}{\begin{eqnarray}}
\newcommand{\ena}{\end{eqnarray}}
\newcommand{\beano}{\begin{eqnarray*}}
\newcommand{\enano}{\end{eqnarray*}}
\newcommand{\sect}[1]{\setcounter{equation}{0}\section{#1}}
\newcommand{\vs}[1]{\rule[- #1 mm]{0mm}{#1 mm}}
\newcommand{\hs}[1]{\hspace{#1 mm}}

\newcommand{\sm}[2]{\frac{\mbox{\footnotesize #1}\vs{-2}}
                   {\vs{-2}\mbox{\footnotesize #2}}}
\newcommand{\nonu}{\nonumber \\}
\newcommand{\half}{\frac{1}{2}}
\newcommand{\shalf}{\sm{1}{2}}
 
\newcommand{\A}{\alpha}

\newcommand{\sig}{\sigma}

\newcommand{\pcr}{\mbox{Poincar\'e }}
\newcommand{\mb}[1]{\hs{4}\mbox{#1}\hs{4}}

\newcommand{\var}{\varphi}

\newcommand{\cz}{\mbox{$\cal{Z}$}}
\newcommand{{\cg}}{\mbox{$\cal{G}$}}
\newcommand{\ch}{\mbox{$\cal{H}$}}

\newcommand{\cs}{\mbox{$\cal{S}$}}

\newcommand{\cw}{\mbox{$\cal{W}$}}

\newcommand{\al}{\alpha}

\newcommand{\prt}{\partial}
\newcommand{\eps}{\epsilon}
\newcommand{\vareps}{\varepsilon}

\newcommand{\cu}{\mbox{${\cal U}$}}

\newcommand{\baral}{\bar{\alpha}} 
 
\newcommand{\NP}[1]{Nucl.\ Phys.\ {\bf #1}}

\newcommand{\CMP}[1]{Comm.\ Math.\ Phys.\ {\bf #1}}

\newcommand{\IJMP}[1]{Int. Journ.\ Mod.\ Phys.\ {\bf #1}}
\newcommand{\PR}[1]{Phys.\ Rev.\ {\bf #1}}

\newcommand{\C}{\mbox{\hspace{.0em}\rule{0.042em}{.674em}\hspace{-.642em} 
\rm C$\,$}}
\newcommand{\R}{\mbox{\hspace{.0em}\rule{.042em}{.694em}\hspace{-.38em}
\rm R$\,$}}

\newcommand{\Z}{\mbox{$Z$\hspace{-1.1em}$Z$ }}

\newcommand{\1}{\mbox{\hspace{.0em}\rule{0.042em}{.7em}\hspace{-.524em}
{\large 1}$\!$}}

\begin{document}
\renewcommand{\thefootnote}{\fnsymbol{footnote}}
\newpage
\pagestyle{empty}
\setcounter{page}{0}
\newcommand{\norm}[1]{{\protect\normalsize{#1}}}
\newcommand{\LAP}
{{\small E}\norm{N}{\large S}{\Large L}{\large A}\norm{P}{\small P}}
\newcommand{\sLAP}{{\scriptsize E}{\footnotesize{N}}{\small S}{\norm L}$
${\small A}{\footnotesize{P}}{\scriptsize P}}
\def\logolapin{
  \raisebox{-1.2cm}{\epsfbox{enslapp.ps}}}
\def\logolight{{\bf{{\large E}{\Large N}{\LARGE S}{\huge L}{\LARGE
        A}{\Large P}{\large P}} }}
\def\logoenslapp{\logolight}
%
%
%
\hbox to \hsize{
\hss
\begin{minipage}{5.2cm}
  \begin{center}
    {\bf Groupe d'Annecy\\ \ \\
      Laboratoire d'Annecy-le-Vieux de Physique des Particules}
  \end{center}
\end{minipage}
\hfill
\logoenslapp
\hfill
\begin{minipage}{4.2cm}
  \begin{center}
    {\bf Groupe de Lyon\\ \ \\
      Ecole Normale Sup\'erieure de Lyon}
  \end{center}
\end{minipage}
\hss}
\vspace{.3cm}
\centerline{\rule{12cm}{.42mm}}

\vfill
 
\begin{center}
 
{\Large {\bf REMARKS ON FINITE $\cw$ ALGEBRAS}}\\[1cm]

{\large{\bf F. Barbarin$^1$, E. Ragoucy$^2$ and P. Sorba$^1$,}}\\[.42cm]
$^1$ {\em Laboratoire de Physique Th\'eorique }\LAP\footnote{URA 14-36
du CNRS, associ\'ee \`a l'Ecole Normale Sup\'erieure de Lyon et \`a
l'Universit\'e de Savoie.}\\
 B.P. 110, F-74941 Annecy-le-Vieux Cedex, France.\\[.24cm]
$^2$ CERN, TH Division, CH-1211 Geneva 23, Switzerland.
 
\end{center}
\vfill
 
\begin{abstract}
 
\indent
 
The property of some finite $\cw$ algebras to be the
commutant of a particular subalgebra of a simple Lie algebra $\cg$
is used to construct realizations of $\cg$.
 
When $\cg \simeq so(4,2)$, unitary representations of the conformal
and Poincar\'e algebras are recognized in this approach, which can
be compared to the usual induced representation technique. When $\cg
\simeq sp(2,\R)$ or $sp(4,\R)$, the anyonic parameter can be seen
as the eigenvalue of  a $\cw$ generator in such
$\cw$ representations of $\cg$.
 
The generalization of such properties to the affine case is also
discussed in the conclusion, where an alternative of the Wakimoto
construction for $\widehat{sl}(2)_k$ is briefly presented.
 
\end{abstract}
 
\vfill
 
This mini review is based on invited talks presented by P.~Sorba at
the {\em ``Vth International Colloquium on Quantum Groups and Integrable
Systems''}, Prague (Czech Republic), June 1996; {\em ``Extended and
Quantum Algebras and their Applications to Physics''}, Tianjin
(China), August 1996; {\em ``Selected Topics of Theoretical and
Modern Mathematical Physics''}, Tbilisi (Georgia), September 1996;
 to be published in the Proceedings.
 
\vfill
\rightline{hep-th/9612070}
\rightline{\LAP-AL-628/96}
\rightline{CERN-TH 96-338}
\rightline{November 1996}
 
\newpage
\pagestyle{plain}
\renewcommand{\thefootnote}{\arabic{footnote}}
\sect{Preliminaries}
 
 The subject of this talk concerns a special class of algebras, which have
been called ``finite $\cw$ algebras'' \cite{1} and which we will denote
FWAs. First constructed from the zero modes of the (known) $\cw$ algebras,
the FWAs  present two
appealing features.
 
First, they can be seen as a good laboratory for studying  properties of the
usual -or affine- $\cw$ algebras, which depend of a complex variable:
this is due to the relative simplicity of their commutation relations,
compared to the
affine case. The
second point is that they constitute their own a rather interesting field of
investigation, as well as applications, in  mathematical physics. It is
this second
aspect of finite $\cw$ algebras  that we wish to raise and develop
hereafter, just mentioning in the conclusion the generalization of
our results to affine algebras.
 
The plan of this review will be the following:
 
-Section 2  contains  a  brief introduction on finite $\cw$
algebras,  with some definitions and notations.
 
-Section 3 deals with the construction of (a class of) finite $\cw$
algebras starting from a simple Lie algebra \cg. In our approach, the FWA
appears as the
commutant, in a generalization of the enveloping algebra of \cg, of
a \cg-subalgebra.  Such an approach  can be seen as providing a definition
of (a class of) FWAs,
and also as an
explicit method for determining the commutant of (a class of) subalgebras of a
simple Lie algebra.
 
Then, it is this property of  FWAs that will be exploited to construct
realizations of a simple Lie algebra $\cg$.  More precisely, it can be
shown that
knowing a special realization of $\cg$ in terms of differential
operators of the $\cg$ generators, new $\cg$ realizations can be
constructed owing to a suitably chosen FWA. This will be the subject
of section 4, where the method is illustrated on the $sl(2,\R)$
algebra. Then:
 
-Section 5 presents such a $\cw$ realization for the four
dimensional conformal algebra $so(4,2)$. Unitary representations of
the conformal algebra, as well as its Poincar\'e subalgebra, can be
recognized in this approach, which can be compared with the usual induced
representation technique.
 
-Section  6  is devoted to $\cw$ realizations of the $sp(2)$
and $sp(4)$ cases. These Lie algebras can be considered  as the
algebras of observables for a system of two identical particles in $d=1$ and
in $d=2$ dimensions, in the Heisenberg quantization scheme.
In each case, it will be possible to relate the anyonic parameter to
the eigenvalues of a $\cw$ generator.
 
Finally we conclude by a short discussion on the extension of these
results to the affine case.
 
\sect{Definitions and notations}
 
\indent
 
 As a  general definition of a finite $\cw$ algebra, we can propose
 the  following :
 
\indent
 
{\it It is an algebra over a field} $k$ {\it (we will limit to}
$\R$ {\it  or}\ {$\C$}) {\it the two corresponding internal laws being
the usual addition and multiplication with an extra internal law ---the
commutator--- which is
antisymmetric,} $k${\it -bilinear, satisfies the Jacobi identity 
and closes polynomially.
 
 We will talk about a classical $\cw$ algebra when the algebra is
endowed with a Kirillov-Poisson structure, and the so-called commutator is
the Poisson bracket (PB); we will speak about a quantum $\cw$ algebra
when the commutator is simply $[A,B] = AB - BA$ for any couple of
elements $A$ and $B$}.
 
\indent
 
With this definition, the enveloping algebra of a finite
dimensional Lie algebra appears as a particular  case of $\cw$
algebra. But let us propose an example:
 
Consider the algebra generated by the four elements  $E,F,H$ and $C$
with the commutation relations (C.R.):
\be
[H,E] = E \;\;\; [H,F]=-F \;\;\; [E,F] =H^2+ C
\label{eq:1}
\ee
\[
\mbox{with} \; [C,E] = [C,F]  = [C,H]=0
\]
 
It seems natural to compare this algebra  with  the $sl(2)$ one
 generated by $e,f,h$:
\be
[h,e] = e  \;\;\;  [h,f]=-f   \;\;\;  [e,f] = 2 h
\label{eq:2}
\ee
 
 The algebra  (\ref{eq:1}) can be obtained from the zero modes of
 the  (affine)  $\cw$ algebra made by four generators, of spin
 2,  3/2, 3/2  and 1 under the spin 2 Virasoro generator,
and sometimes
 called the Bershadsky algebra  $\cw^{(2)}_3$.The zero
mode of the spin 2 element is $C$, the ones of spin 3/2  are  $E$
and $F$, while the one of spin 1 is $H$. One must note that
commuting the zero modes of two different spin 3/2 generators  does
not uniquely provide zero modes in the other generators.
The standard  procedure to eliminate the non-zero modes
consists in considering the action  of such
elements on any highest-weight representation of the $\cw$ algebra;
then projecting out both sides of the C.R. on each h.w. state will
allow us to get the C.R. of  (\ref{eq:1}), the positive modes
annihilating the h.w. states.
 
The $\cw^{(2)}_3$ algebra belongs to the large class of $\cw$
algebras, symmetries of Toda theories, and which can be constructed from WZW
models by using the Hamiltonian reduction technique\cite{2}. The first
step consists in imposing (first-class) constraints on the components of the
conserved currents in the considered  WZW model. Such constraints imply gauge
transformations, and the associated $\cw$ algebra will then be
obtained by determining the corresponding gauge-invariant polynomial
quantities.
 
In the following, $\cg$ will stand for the Lie algebra of a simple,
real, connected and non-compact Lie group $G$. Let $\{ t_a \}, a=1
... dim \cg$, be a basis of $\cg$ and $\{ J^a \}$ the dual basis in
$\cg^*$
\be
[t_a, t_b] = f^c_{ab} t_c \;\;\; J^a(t_b) = \delta^a_b
\label{eq:2.3}
\ee
We introduce the metric on $\cg$ in a representation $R$:
\be
\eta_{ab} = <t_a,t_b> = tr_R(t_at_b) \; \mbox{and} \;  \eta_{ab}
\eta^{bc} = \delta_a^c
\label{eq:2.4}
\ee
 
We can then define on $\cg^*$ a Poisson-Kirillov structure that
mimicks the commutator (we have identified $\cg^*$ and $\cg$):
\be
\{J^a,J^b\} = f^{ab}_c J^c \; \; \mbox{with} \; \; f^{ab}_c
= \eta^{ad} \eta^{be} \eta_{cg} f^g_{de}
\label{eq:2.5}
\ee
 
Actually, from the simple Lie algebra \cg, the different sets of
constraints one wishes to impose, and therefore the different $\cw$
algebras one can construct, are in one-to-one correspondence with the
embeddings of the $sl(2)$ algebra in
$\cg$. To each such embedding  can be associated a grading of
 $\cg$, given by the eigenvalue of the $sl(2)$ Cartan generator $h$:
$\cg = \oplus^{+m}_{p=-m}  \cg$ with $[h,X] = pX$ for any $X \in \cg ,
p \in \shalf\ \Z$; we then have the property 
$[\cg_p, \cg_q] \subset \cg_{p+q}$,
with the convention that $\cg_p = \{0\}$ when $|p|>m$.  More generally, we
write:
\be
\cg = \cg_+ \oplus \cg_0 \oplus \cg_-
\ee
the constraints being imposed, in the current matrix  $J = J^a t_a$,
on the components of $\cg_-$.
 
Since any $sl(2)$ subalgebra in a classical simple $\cg$ algebra is
principal in a subalgebra $\ch$ of $\cg$ (we discard the exceptional
algebras case), it is quite usual to denote the corresponding $\cw$
algebra by $\cw(\cg, \ch)$. Taking as an example the $\cg = sl(3)$
case, two $\cw$ algebras can be constructed in this way: the $\cw
(sl(3),sl(3))$ one, which in the affine case is generated by the two
fields $W_2$ and $W_3$ of respective spin 2 and 3, usually also
written as $\cw^{(1)}_3$, and the $\cw (sl(3), sl(2))$ one, 
presented just above as the
Bershadsky $\cw$ algebra and also denoted as $\cw^{(2)}_3$.
 
\sect{Realization of finite $\cw$ algebras}
 
In the usual Hamiltonian reduction approach \cite{2}, we start by
imposing (first-class) constraints on the $\cg_-^*$ part of the $J$-matrix.
Following Dirac's prescription, these first-class constraints generate
a gauge invariance on the $J^a$'s, i.e. in the classical case:
\be
J \rightarrow J^g = exp \left( c_\al \{ J^\al , \cdot \}_{cons.}
\right) (J) \ee
where the $\{ \; , \; \}_{cons.}$ means that one has to impose the
constraint conditions on the r.h.s. of the Poisson bracket.
Developing $J^g$ with the help of the gradation and the use of
constraints, we can, using the relations (\ref{eq:2.3}-\ref{eq:2.5}),
rewrite $J^g$ as:  \be
J^g = exp \left( c^{\bar{\al}} [t_{\baral}, \cdot] \right) (J) =
g_+^{-1} Jg_+ \label{eq:3}
\ee
where $g_+ = exp (c^{\baral} t_{\baral})$ with the parameters
$c^{\baral} = \eta^{\baral \al} c_\al$ and $t_{\baral} \in \cg_+$.
Thus the gauge transformations can be seen as conjugation on $\cg$
by elements of the subgroup $G_+$.
 
Finally, while fixing the gauge, one obtains, in the components of
$J^g$, gauge-invariant quantities, i.e. quantities which
Poisson-commute with the constraints.
 
The main idea of our construction \cite{3,4} is not to impose constraints
anymore on $\cg^*_-$
once  the gradation is chosen, but however to use
the $G_+$ conjugation as in (\ref{eq:3}):
\be
J^g_{tot} = g^{-1}_+ J_{tot} \; \; g_+ \; \; \mbox{with} \; \; g_+
\in G_+ \ee
(we denote the $J$-part by $J_{tot}$ in order to emphasize that no
restrictions on its components have been put).
 
Then, developing $J^g_{tot}$ by using the same rules as before, we get:
\be
J^g_{tot} = exp \left( c_\al \{J^\al, \cdot \} \right)  (J_{tot})
\ee
where now the PBs are computed without using what were the
constraints. Thus if one finds quantities which are invariant under
the coadjoint transformations, these objects will have strongly
vanishing PBs with the elements $J^\al$. Let us add that, although
the transformations we are looking at have the same form as the
 gauge transformations described 
at the beginning of this section, they are not
gauge transformations, not being associated with constraints. Thus the
construction we present is strictly algebraic, but we will see that
the technique can be applied to physical problems.
 
Then we are looking for quantities which Poisson-commute with the
$\cg_-^*$ part of $\cg^*$. Let us translate this problem into the
Lie algebra $\cg$. Such a quantization can be easily performed
thanks to the Lie isomorphism between $\cg^*$ and $\cg$, and
a symmetrization procedure that maps polynomials in $\cg^*$ onto
elements of $\cu(\cg)$. Indeed, the isomorphism $i$ between $\cg^*$
and $\cg$ defined by: \be
i(J^a) = t^a = \eta^{ab} t_b
\ee
where $\eta^{ab}$ is the inverse matrix of the metric $\eta_{ab}$,
can be extended as a vector space homomorphism from $\cg^*$
polynomials into $\cu(\cg)$ with the rule:
\be
i(J^{a_1} J^{a_2} \cdots J^{a_n}) = S (t^{a_1} \cdot t^{a_2}
\cdots t^{a_n}) \; \; \; \; \forall n
\ee
where $S(\cdot , \cdots , \cdot)$ stands for the
symmetrized product of the generators $t^a$; $S$ is normalized by
$S(X,\cdots , X)=X^n$.
 
At this stage, one could realize that the finite $\cw$ algebras that
we wish to construct have some connection with the commutant of a
subalgebra in $\cg$. Actually, after developing a symmetry fixing
procedure, and limiting to a class of finite $\cw$ algebras that
are of special interest for the rest of this talk, we can announce
the following result \cite{4}:
 
\indent
 
\underline{Theorem:} {\it Any finite $\cw(\cg, \cs)$ algebra, with
$\cs= \mu sl(2)$ regular subalgebra of $\cg$, can be seen as the
commutant in a (localization of) the enveloping algebra $\cu (\cg)$ 
of some $\cg$-subalgebra $\tilde{\cg}$.
 
Moreover, let $H$ be the Cartan generator of the diagonal $sl(2)$
in $\cs$. If we call $\cg_-, \cg_0$ and $\cg_+$ the eigenspaces of
respectively negative, null and positive eigenvalues under $H$, then
$\tilde{\cg}$ decomposes as $\tilde{\cg} = \cg_- \oplus
\tilde{\cg}_0$, where $\tilde{\cg}_0$ is a subalgebra of $\cg_0$,
which can be uniquely determined}.
 
\indent
 
Let us briefly comment this property. First of all, the $\cg$
grading obtained from regular subalgebras $\cs = \mu sl(2)$ are
always such that the $\cg_-$ part is Abelian. This means in particular
that, in the determination of quantities that commute with $\cg_-$,
the elements of $\cg_-$ themselves appear. In order to get rid of
these undesirable quantities, one can think about increasing the
subalgebra $\cg_-$ up to another one $\tilde{\cg}$, in such a way
that the commutant of $\tilde{\cg}$ provides exactly the $\cw$
algebra one wishes to obtain.
 
Note that more general finite $\cw(\cg, \ch)$ algebras than the ones
mentioned above can be obtained as a commutant, but one will then
have to extend $\tilde{\cg}_0$ to a part of $\cg_+$ in $\cg$. The
case of ``affine'' $\cw(\cg,\ch)$ algebra can also be treated within
this framework.
 
Thus the $\cw$ algebras that one can construct are written in terms
of all the generators of the algebra $\cg$. There is, however, a price to
pay: the generators obtained this way show up as functions
$P(t_a)/Q(t_\al)$ with $P$ a polynomial in all the $t_a$'s and $Q$
a smooth function in the center of the $\widetilde{\cg}$ Lie
derivative (i.e. $\cz ([\widetilde{\cg}, \widetilde{\cg}])$).
It can be shown that the $\cw$ generators form a polynomial basis of
the commutant of $\widetilde{\cg}$ in a generalization of the
enveloping algebra $\cu(\cg)$. The one we consider is the
localization $\cu(\cg)_{\cal S}$, where 
$\cs =\cz ([\widetilde{\cg}, \widetilde{\cg}])$,
 which contains apart from $\cu(\cg)$ itself, quotients
$u^{-1}v, v u^{-1}$, where $u \in \cs, u \neq 0$ and $v \in \cu(\cg)$,
or an extension of this latter allowing elements like $u^r, u \in \cs,
r \in \half\ \Z$.
Let us emphasize that the technique that is summarized and briefly
commented on above leads to a purely algebraic construction of a class
of finite, as well as affine, $\cw$ algebras. Consequently, it can
also be considered as a way of defining (a family of) $\cw$ algebras.
 
\sect{$\cw$ realizations of simple Lie algebras}
 
\indent
 
The property of a $\cw$ algebra to appear as the commutant of a
$\cg$-subalgebra can be used to build, from a special realization of
$\cg$, a large set of $\cg$-representations \cite{3}\cite{5}. Indeed
one knows how to construct a realization of $\cg$ with differential 
operators on
the space of smooth functions $\varphi (x_1, \cdots,x_n)$ with
$n=dim \cg_-$. In this picture, when $\cg_- \equiv \cg_{-1}$, the
abelianity of the $\cg_{-1}$ part allows each $\cg_{-1}$ generator to act
by direct multiplication:
\be
\varphi (x_1, \cdots, x_n) \rightarrow x_i\varphi (x_1, \cdots, x_n)
\; \; \; \; \mbox{with} \; \; \; \; i=1, \cdots, n
\label{eq:4.1}
\ee
---cf. action of the translation group--- while the generators of the
$\cg_0 \oplus \cg_+$ part will be represented by polynomials in the
$x_i$ and $\prt_{x_i}$.
 
It is from a particular ---canonical--- differential realization of
$\cg$ that new realizations will be constructed with the use of
the finite $\cw$ algebra mentioned above. Realization of the
$\tilde{\cg}$ generators will not be affected in this approach. On
the contrary, to the differential form of each generator in a
certain supplementary subspace of $\widetilde{\cg}$ in $\cg$ will be
added a sum of $\cw$ generators, the coefficients of which
 $f(x_i, \prt_{x_i})$ are polynomials in the $\prt_{x_i}$'s.
To each irreducible $d$-dimensional representation of the $\cw$
algebra one can associate a matrix differential realization of $\cg$
acting on vector functions $\bar{\var} = (\var_1, \cdots, \var_d)$
with $\var_i = \var_i(x_1, \cdots, x_n)$.
 
It is time to illustrate our technique on the simplest non-trivial
example, i.e. $\cg = sl(2,\R)$.
 
Let us define:
\be
J = J^-  t_- + J^0 t_0 + J^+ t_+ =
\left(
\begin{array}{cc}
J^0 & J^+ \\
J^- & -J^0
\end{array}
\right)
\label{eq:4.2}
\ee
with $\cg_{+,0,-}$ generated by $t_{+,0,-}$ respectively.
 
By the action of an adequate $G_+$ element, namely:
\be
g_+ =
\left(
\begin{array}{cc}
1 & -J^0/J^- \\
0&1
\end{array}
\right)
\label{eq:4.3}
\ee
one obtains (symmetry fixing):
\be
J^{g_+} = g_+  \; J g^{-1}_+ =
\left(
\begin{array}{cc}
0 & \frac{J^+ J^- + (J^0)^2}{J^-} \\
J^- & 0
\end{array}
\right)
\label{eq:4.4}
\ee
 
It follows, after quantization:
\be
J^+ J^- + (J^0)^2 \rightarrow \shalf (t_+ t_- + t_- t_+) + t_0^2 =
C_2
\label{eq:4.5}
\ee
that is, exactly the $C_2$ Casimir operator of $sl(2,\R)$ generating
the finite $\cw$ algebra that we wish to determine (do not forget
that we are expecting the zero mode of the Virasoro generator~!).
 
On this simple example, we can convince ourselves that the commutant
of $\cg_-$ in $\cu(\cg)_{\cal S}$ ($\cs$ is generated by $t_-$) denoted by
$Com (\cg_-)$ is a polynomial algebra generated by $\{ C_2,
t_- , \frac{1}{t_-}\}$.
 
In order to get the $C_2$ element only, we will look for the
commutant of a Lie algebra larger than $\cg_-$; more precisely, we
will obtain:
\be
Com \; (\cg_- \oplus \cg_0) = Polyn \ (\{ C_2 \}).
\label{eq:4.7}
\ee
 
Now, let us show how this construction can be applied to
realizations of $sl(2,\R)$. For such a purpose, consider
the $sl(2)$ (differential) realization:
\be
E_- = \shalf x^2 \; \; \; \; E_+ = - \shalf \prt^2_x \; \; \; \;
H=-(x^2 \prt_{x^2} + \sm{1}{4} )
\label{eq:4.8}
\ee
acting on smooth functions $\var$ of the real variable $x$. We are
in the conditions of eq. (\ref{eq:4.1}) for the $\cg_{-1}$ part, which
is one-dimensional. We also note that the eigenvalue of $C_2$ for
this representation is $=-3/16$.
 
On this example, it is an easy calculation to write down from
(\ref{eq:4.8}) new realizations of the algebra under consideration.
Leaving the $\cg_-$ generator unchanged, as well as the $\cg_0$ one,
a realization corresponding to the eigenvalue $\gamma$ of $C_2$ is
given by:
\be
E_- = \shalf x^2 \; \; \; \; \; \; \; \; E_+ = - \shalf \prt^2_x +
\frac{\gamma
+ \sm{3}{16}}{x^2} \; \; \; \; \; \; \; \; H = -(x^2 \prt_{x^2} + \sm{1}{4})
\label{eq:4.9}
\ee
 
The above expression of $E_+$ can also be obtained systematically in
the following way (which, therefore, generalizes to the other
cases). Coming back to eq. (\ref{eq:4.4}), let us formally act on
$J^{g_+}$ by $g^{-1}_+$:
\be
J^{g_+} =
\left(
\begin{array}{cc}
0& \frac{"C_2"}{J^-} \\
J^- & 0
\end{array}
\right)
\rightarrow
g^{-1}_+ J^{g_+} g_+ = J =
\left(
\begin{array}{cc}
J^0 & \frac{-(J^0)^2 + "C_2"}{J^-} \\
J^- & -J^0
\end{array}
\right)
\label{eq:4.10}
\ee
where $"C_2"$ denotes the Casimir element before quantization. By
identification, one gets:
\be
J^+ = - \frac{(J^0)^2}{J^-} + \frac{"C_2"}{J^-}
\label{eq:4.11}
\ee
which shows a direct correspondence between $J^+$ and $"C_2"$. $J^+$
is linear in $"C_2"$ and vice-versa; this linearity property will
survive at the quantum level:
\be
E_+ = \frac{1}{E_-} (C_2 - H^2 - H)
\ee
which leads to the $E_+$ expression in (\ref{eq:4.9}).
 
Thus, from the special (``canonical'') differential realization
(\ref{eq:4.8}), our technique on the commutant has allowed to
get a large class of $sl(2,\R)$ representations.
 
\sect{Unitary irreducible representations of the conformal and
Poincar\'e algebras}
 
\indent
 
We now apply the results summarized in sections 3 and 4 to get $\cw$
realizations of the $so(4,2)$ algebra and of its Poincar\'e
subalgebra \cite{5}. The $so(4,2)$ algebra is known as the conformal
algebra in four dimensions in the Minkowski space, i.e. with the
metric $g_{\mu \nu} = diag(1,-1,-1,-1)$. Its fifteen generators can be
chosen and realized in the momentum representation as follows:
 
- four
translations: $P_\mu = p_\mu$ $(\mu =0,1,2,3)$ forming the $\cg
_-$ part.
 
- six Lorentz generators: $M_{\mu \nu} = i(p_\mu \prt_\nu - p_\nu
\prt_\mu)$ forming with the dilatation $D=-i(p\cdot \prt +4)$ the
$\cg_0$ part.
 
-  four special conformal transformations
\be
K_\mu = p_\mu \Box -2p- \prt \cdot  \prt_\mu -8 \prt_\mu
\ee
constituting the $\cg_+$ part.
 
The corresponding grading operator is $D$, and in the above
expressions $\prt_\mu$ stands for $\prt / \prt p^\mu ,p$ is the
quadrivector $(p_\mu) , \prt$ the quadrivector $(\prt_\mu)$ and
$\Box = \prt \cdot \prt = g^{\mu \nu} \prt_\mu \prt_\nu = \prt^\mu \prt_\mu$.
 
We wish to construct the commutant of the $P_\mu$ in order to build
a $\cw$ realization of $\cg = so(4,2)$. Note that $so(4,2)$ and
$sl(4,\R)$ are two different non compact real forms of the algebra
$so(6) \sim su(4)$. If we had considered $sl(4,\R)$, i.e. the maximally 
non-compact form of $\cg$, the chosen gradation would correspond to the
model $\cw(sl(4), 2sl(2))$. Referring to section 3, we can take
$\tilde{\cg} = \cg_- \oplus \tilde{\cg}_0$, where:
\be
\tilde{\cg} = \{ M_{13} - M_{01}, M_{23} - M_{02}, M_{03}, D \}
\ee
 The commutant of $\tilde{\cg}$ can therefore be seen as a
compactified form of $\cw (sl(4),$ $2sl(2))$. It contains seven
generators: three generators $J_k, k = 1,2,3$, forming an $so(3)$
algebra; three other generators $S_l, l=1,2,3$ forming a vector
under this $so(3)$, their C.R.'s  closing under a polynomial
in the $J_k$ and the seventh generator $C_2$, which is the 
second-order Casimir of $so(4,2)$. In summary the $W$ C.R.'s read: 
\bea
{[} J_j , J_k ] &=& i \ \vareps_{jkl} J_l \nonumber \\
{[} J_j , S_k ] &=& i \ \vareps_{jkl} S_l \nonumber \\
{[} S_j, S_k ] &=& -i \ \vareps_{jkl} (2(J^2_1 + J^2_2 + J^2_3) -C_2
-4) J_l \nonumber \\
{[} C_2 , J_j ] &=& [C_2, S_j ] =0 \; \; \ \; \ \; \{ j,k,l \}
=\{ 1,2,3 \}.
\label{eq:5.3}\ena
 
It can  directly be checked that this algebra satisfies, for each (real)
value of the $C_2$ scalar, the defining C.R.'s of the
Yangian \cite{6bis}
$Y(sl(2))$, with generators $J_i$ and $S_i (i=1,2,3)$. In other
words, the $\cw$ algebra defined by (\ref{eq:5.3}) provides a
realization of\footnote{We thank Mo-Lin Ge for bringing our
attention to this point.} $Y(sl(2))$.
 
The $so(4,2)$ realization obtained by our $\cw$ approach
stands as follows, for $p^2 >0$ (note that sign ($p^2)$ = $+,0,-$ is
conserved in $so(4,2)$):
\bea
P_\mu &=& p_\mu\, \1 \mb{ }\mu=0,1,2,3 \label{Pmu}  \\
M_{12} &=& i (p_{1}\prt_{2} - p_{2}\prt_{1} )\, \1 +J_3 \label{M12}  \\
M_{13}&=& i (p_{1}\prt_{3} - p_{3}\prt_{1} ) \, \1
-\frac{\sqrt{p^2}}{p_0+p_3}J_2 -\frac{p_2}{p_0+p_3}J_3 \\
M_{23} &=& i (p_{2}\prt_{3} - p_{3}\prt_{2} )\, \1
+\frac{\sqrt{p^2}}{p_0+p_3}J_1 +\frac{p_1}{p_0+p_3}J_3 \\
M_{01} &=& i (p_{0}\prt_{1} - p_{1}\prt_{0} ) \, \1
-\frac{\sqrt{p^2}}{p_0+p_3}J_2 -\frac{p_2}{p_0+p_3}J_3 \\
M_{02} &=& i (p_{0}\prt_{2} - p_{2}\prt_{0} )\, \1
+\frac{\sqrt{p^2}}{p_0+p_3}J_1 +\frac{p_1}{p_0+p_3}J_3 \\
M_{03} &=& i (p_{0}\prt_{3} - p_{3}\prt_{0} ) \, \1  \label{M03} \\
D &=& -i ( p\cdot\prt +4)\, \1 \label{dilat}   \\
K_0 &=& (p_{0}\Box -2 p\cdot\prt\prt_{0}  - 8\prt_0 )\, \1
-\frac{2}{p_0+p_3}Z_3 +\frac{p_0}{p^2}Z_0 \nonu
&&+\frac{1}{(p_0+p_3)\sqrt{p^2}}(p_1 Z_1 +p_2 Z_2) +\nonu
&& -\frac{2i \sqrt{p^2}}{p_0+p_3}
\left( -(\frac{5}{2} \frac{p_2}{p^2} +\prt_2 )J_1
+(\frac{5}{2} \frac{p_1}{p^2} +\prt_1)J_2 \right)\nonu
&&- \frac{2i}{p_0+p_3}(p_2\prt_1-p_1\prt_2)J_3 \label{K0}   \\
K_1 &=& (p_{1}\Box -2 p\cdot\prt\prt_{1}  - 8\prt_1 )\, \1
+\frac{1}{\sqrt{p^2}}Z_1 +\frac{p_1}{p^2}Z_0 \nonu
&& -\frac{2i \sqrt{p^2}}{p_0+p_3}
\left( \frac{5}{2} \frac{p_0+p_3}{p^2} +\prt_0 +\prt_3 \right)
J_2 \nonu
&&-2i \left( \frac{p_2}{p_0+p_3}(\prt_0+\prt_3)-\prt_2 \right)J_3 \\
K_2 &=& (p_{2}\Box -2 p\cdot\prt\prt_{2}  - 8\prt_2 )\, \1
+\frac{1}{\sqrt{p^2}}Z_2 +\frac{p_2}{p^2}Z_0 \nonu
&& +\frac{2i \sqrt{p^2}}{p_0+p_3}
\left( \frac{5}{2} \frac{p_0+p_3}{p^2} +\prt_0 +\prt_3 \right)
J_1  \nonu
&& +2i \left( \frac{p_1}{p_0+p_3}(\prt_0+\prt_3)-\prt_1 \right)J_3 \\
K_3 &=& (p_{3}\Box -2 p\cdot\prt\prt_{3}  - 8\prt_3 )\, \1
+\frac{2}{p_0+p_3}Z_3 +\frac{p_3}{p^2}Z_0  \nonu
&& -\frac{1}{(p_0+p_3)\sqrt{p^2}}(p_1 Z_1 +p_2 Z_2) \nonu
&&+\frac{2i \sqrt{p^2}}{p_0+p_3}
\left( -(\frac{5}{2} \frac{p_2}{p^2} +\prt_2 )J_1
+(\frac{5}{2} \frac{p_1}{p^2} +\prt_1)J_2 \right)  \nonu
&& + \frac{2i}{p_0+p_3}(p_2\prt_1-p_1\prt_2)J_3 \label{K3}
\ena
where
\be
\begin{array}{ll}
Z_1=2S_1+J_3J_1+J_1J_3 \ \ &  Z_2=2S_2+J_3J_2+J_2J_3  \\
\; & \; \\
Z_3=S_3-(J_1^2+J_2^2)  & Z_0=2S_3+C_2-J_3^2-2(J_1^2+J_2^2)
\label{basis}
\end{array}
\ee
 
Let us focus for the moment on the expressions of the Poincar\'e
generators and remark that only the $\vec{J}$-part of the $\cw$
algebra shows up there. We recall the expressions of the
Pauli-Lubanski-Wigner quadrivector
$W^\mu=\half\eps^{\mu\nu\rho\sig} P_\nu M_{\rho\sig}$, which satisfies:
\be
{[W_\mu, P_\nu]}=0 \mb{ }
{[W_\mu, W_\nu]}=i\eps_{\mu\nu\rho\sig} W^\rho P^\sig \mb{ }
 W\cdot P=0
\ee
It is well-known that the irreducible representations of the \pcr algebra are
labelled by the eigenvalues of $P^2=p^2$ and $W^2=-s(s+1)p^2$,
where $s$ is the spin of the particle. Because of the relation
$W\cdot P=0$, the quadrivector $W^\mu$ possesses only 3 independent
components. These generate the spin algebra
$so(3)$ when $p^2$ is positive.
This is recovered in a very natural way in our \cw\ algebra framework.
Indeed, the generators $J_k$ really play the role of the spin generators, since
they can be rewritten as:
\be
J_k=-\frac{1}{m} n_k\cdot W=- \frac{1}{m} (n_k)^\mu W_\mu
\mb{and} W^\mu=-m\sum_{k=1}^{k=3} (n_k)^\mu J_k
 \label{defJ}
 \ee
(since $P\cdot W=0$)
\be
\mbox{with} \qquad m=\sqrt{p^2} \qquad \mbox{and} \qquad
W \cdot W = -P^2 \vec{J}^2 \label{WW}
\ee
and where we have introduced
 the frame \cite{6} of the ``particle'' of momentum $p$:
\bea
n_0 &=& (n_0)_{\mu} = \frac{1}{m}\, p\ =\
(\frac{p_0}{m},\frac{p_1}{m},\frac{p_2}{m},\frac{p_3}{m})\nonu
n_1 &=& (n_1)_{\mu} = \frac{1}{p_0+p_3}(p_1,p_0+p_3,0,-p_1)
\nonu
n_2 &=& (n_2)_{\mu} = \frac{1}{p_0+p_3}(p_2,0,p_0+p_3,-p_2)
\nonu
n_3 &=& (n_3)_{\mu} = \frac{-m}{p_0+p_3}(1,0,0,-1) +\frac{1}{m}\, p
\label{eq:n3}
\ena
which obeys $n_\mu\cdot n_\nu=(n_\mu)^\rho (n_\nu)^\sig
g_{\rho\sig}=g_{\mu\nu}$ and also $(n_\mu)^\rho (n_\nu)^\sig
g^{\mu\nu}= g^{\rho\sig}$.
 
The Lorentz transformation $L(p)$, which moves the rigid referential
frame $(e_0, e_1,$ $e_2, e_3)$ with $e_0 = (1,\vec{0}) , e_1 =
(0,1,0,0) , e_2 = (0,0,1,0)$ and $e_3=(0,0,0,1)$ to the $p$-frame
$(n_0, n_1, n_2, n_3)$, also relates the three-vector $(J_i)$ to the
four-vector $(\cw_\mu)$
\be
mJ = (0, m \vec{J}) \stackrel{L(p)}{\longrightarrow} W = (W
_\mu) \ee
 It is through $L(p)$
that the representations of the Poincar\'e group can be constructed from
representations of the rotation subgroup. Indeed, the Lorentz
transformation $\Lambda$ acting on functions
$\tilde{\varphi}$
of the $p$-variable in the $U$ representation:
\be
^{\Lambda}\tilde{\varphi}(p)=U(\Lambda)\tilde{\varphi}(\Lambda^{-1}p)
\ee
is written more conveniently on the Wigner functions $\psi$ defined
by
\be
\psi(p)=U(L(p)^{-1})\tilde{\varphi}(p)
\ee
as
\be
\left( ^{\Lambda}\psi \right) (p)=U\left( L(p)^{-1}\Lambda
L(\Lambda^{-1}p) \right) \psi(\Lambda^{-1}p).
\label{eq:***}
\ee
We recognize in the product $L(p)^{-1}\Lambda L(\Lambda^{-1}p)$ a Wigner
rotation, element of the $(m,0,0,0)$-vector stabilizer, itself isomorphic to
the $SO(3)$ group when $p^2>0$. It is exactly the infinitesimal
part of  (\ref{eq:***}) that we have in the expressions of the Poincar\'e
generators displayed above.
 
We now look at the other generators of the conformal algebra. In the
same way as we have introduced the
Pauli-Lubanski-Wigner vector $W^\mu$, let us define:
\bea
\Sigma_\mu &=& -W^2\, P_\mu +
P^2 [ \ P^\A M_{\A\mu}(D+i)- \nonu
&& - \half(P_\mu\, P\!\cdot\! K-
P^2\, K_\mu) -\half\eps_{\mu\nu\rho\sig} W^\nu M^{\rho\sig}
] \label{defSig}
\ena
It satisfies in particular
\be
{[\Sigma_\mu, P_\nu]}=0 \; \; \; \; \mbox{and} \; \; \; \;
\Sigma\cdot P=0
\ee
and we can prove that the generators $S_i$ are connected to the
quadrivector $(\Sigma^\mu)$ through
\be
S_k=-\frac{1}{m^3}n_k\cdot \Sigma=-\frac{1}{m^3} (n_k)^\mu \Sigma_\mu
\mb{and} \Sigma^\mu=-m^3\,\sum_{k=1}^{k=3} (n_k)^\mu S_k
\label{defS}
\ee
(since $P\cdot \Sigma=0$)
and also:
\be
m^3S=(0,m^3\vec{S})\ \stackrel{L(p)}{\longrightarrow}\
\Sigma=(\Sigma_\mu)
\ee
 
It would be interesting to add to these geometrical properties a
physical meaning for $(\Sigma_\mu)$, which appears as a sort of
conformal analogous of $(W_\mu)$.
 
Finally, as for the Poincar\'e subcase, we expect that the obtained
$W$-realizations of the $so(4,2)$ algebra can be compared with the
ones constructed via the induced representation method. This later
approach can be found in ref. \cite{7}, where the classification of
all the unitary ray representations of the $su(2,2)$ group with
positive energy is achieved. A direct comparison with the
construction of ref. \cite{7} can be performed \cite{5}, which leads
to a selection of finite-dimensional  representations of the $\cw$
algebra leading to the unitary conformal representations. In ref. \cite{7} the
induced representations are labelled by two non-negative (half)
integers $(j_1, j_2)$ associated with spinor representations of the
Lorentz group $D^{j_1, j_2}$, and by $d$ a real number associated to
the dilatation. In particular the representations of positive masses
satisfy the conditions:
\be
d \geq j_1 + j_2 +2 \; \; \;  \mbox{{with}} \; \; \;
 j_1, j_2 \neq 0 \; \; \;    \mbox{{and}} \;  \; \;
 d > j_1 + j_2 +1 \; \; \;    \mbox{{with}}  \; \; \;
   j_1 , j_2 =0 \label{eq:XX}
\ee
 
Our task is greatly facilitated by the Miura transformation, which
allows the $\cw$ generators to be expressed in terms of generators of the
$\cg_0$ part, that is the Lorentz algebra generated by the rotations
$\vec{R}$ and the boosts $\vec{B}$ to which has to be added the
dilatation $D$. The result is quite simple:
\be
\vec{J} = \vec{R} \ \ \  \ \ \ \vec{S} = \vec{R} \times \vec{B} - i (D-1)
\vec{B}  \ \ \ \ \ \  C_2 = \vec{R}^2 - \vec{B}^2 + D (D-4)
\label{eq:XXX}
\ee
 
Owing to (\ref{eq:XXX}) we know how to associate, and explicitly
construct, the $\cw$ representation relative to the $so(4,2)$ unitary
representation labelled by $(j_1,j_2;d)$ the condition (\ref{eq:XX})
becoming:
\be
\begin{array}{lll}
c_2 \geq 2j_1 (j_1+1) + 2j_2 (j_2+1) + (j_2+j_2)^2 -4 & \mbox{if} &
j_1j_2 \neq 0 \ \ \ \mbox{and} \\
c_2 > 3(j_1 +j_2 +1) (j_1 + j_2-1) &  \mbox{if} & j_1 j_2=0
\end{array}
\ee
where $c_2$ is the $C_2$ eigenvalue.
 
\sect{Anyons and $\cw$ algebras}
 
\indent
 
We now turn to the Heisenberg quantization for a system of
two identical particles in $d=1$ and $d=2$ dimensions\cite{8}. In each case a
finite $\cw$ algebra will be recognized \cite{3} from the algebra of
observables, and used for an algebraic treatment of intermediate
statistics.
 
\subsection{Two particles in $d=1$}
 
We must remark that such a one dimensional system of two identical
particles has been proposed as anyon candidate. Indeed it can be
formally related to a system of two identical vortices in a thin,
incompressible superfluid film, the two spatial coordinates of the
vortex center acting as canonically conjugate quantities \cite{9}.
 
The relative coordinate and momentum of the two particle system are
denoted by:
\be
x = x_{(1)} - x_{(2)} \; \; \; \; \; \; \; p= \shalf (p_{(1)} -
p_{(2)})
\ee
and satisfy the C.R. $[x,p] =i$ in the quantum case.
 
Then the chosen observables, i.e. the quadratic polynomials homogeneous in $x$
and $p$ ($x^2, p^2, xp + px$) close in the quantum case
under the C.R.'s of $\cg= sp(2,\R) \simeq sl(2,\R)$ and we recognize
the expressions already written in (\ref{eq:4.8}). The
$\cw$ treatment \cite{3}
leads directly to (\ref{eq:4.9}), in particular to the expression:
\be
E_+ = \; - \shalf \prt^2_x \; + \; \frac{\gamma + 3/16}{x^2}
\ee
where we can recognize $(x= x_{(1)} - x_{(2)})$ the Calogero
Hamiltonian. As discussed in \cite{8} the parameter $\lambda =
\gamma + 3/16$ can be directly related to the anyonic continuous
parameter, with end point $\lambda =0$ or $\gamma = -3/16$
corresponding to the boson and fermion cases.
 
\subsection{Two particles in $d=2$}
 
Then, the algebra of observables is generated by the quadratic
homogeneous polynomials in the relative coordinates $x_j$ and $p_j\ (j=1,2)$.
One gets a realization of the $\cg=sp(4,\R)$ Lie algebra, the
generators of which can be conveniently separated into three subsets:
 
- the $\cg_{-1}$ part with the three (commuting) coordinate
operators: \be
u = (x_1)^2 + (x_2)^2 \; \; \; \; \; \;  v = (x_1)^2 -(x_2)^2 \; \; \; \;
\; \;  w=
2x_1x_2 \label{eq:10}
\ee
 
- the $\cg_{+1}$ part with the three (commuting) second order
differential operators: \be
U = (p_1)^2 + (p_2)^2 \ \ \ \ \ \; \; \; \; \; \;  V=(p_1)^2 -
(p_2)^2 \; \; \; \; \; \; \ \ \ \ \ W=2p_1p_2
\label{eq:11}
\ee
 
- and the $\cg_0$ part isomorphic to $s\ell(2,\R) \oplus
{g}\ell(1)$  with the four first order
differential operators:
\be
\begin{array}{lll}
C_s = \frac{1}{4} \sum_{i=1}^2 (x_i p_i + p_i x_i) & \; &C_d =
\frac{1}{4} (x_1 p_1 + p_1 x_1 - x_2 p_2 - p_2 x_2) \\
\; & \; \\
L=x_1p_2 - x_2p_1 & \; & M= x_1 p_2 + x_2 p_1
\end{array}
\label{eq:12}
\ee
$C_s$ being the Abelian factor.

The finite $\cw$ algebra associated with this $\cg$-gradation is 
four-dimensional, and can be seen as a ``deformed'' $gl(2)$ algebra, i.e.:
\bea
{[} S, Q ] &=& -2i R \nonumber \\
{[} S, R ] &=& -2i Q \nonumber \\
{[} Q, R ] &=& -8i S (\mu - 2 S^2)  \\
{[} \mu , Q ] &=& [\mu, R ] = [\mu, S] =0\nonumber
\ena
 
We will not give explicit $\cw$ realizations of the $sp(4,\R)$
algebra in order not to overload the text, but rather concentrate
our attention on the possible  determination of an operator which
carries the intermediate statistics in this framework.
 
By the following change of variables
\be
u = r^2 \; \; \; \; \; \;  v=r^2 \sin \theta\; \cos 2 \phi
\; \; \; \; \; \;  w = r^2 \sin \theta \; \sin 2 \phi
\ee
with $0 \leq \theta \leq \frac{\pi}{2}, 0 \leq \phi \leq M,
\mbox{the generator $L$ in}$ (\ref{eq:12}) becomes:
\be
L= -i \frac{\prt}{\prt \phi}
\label{eq:6.8}
\ee
 
We note that when working with univalued functions, this operator
is not well defined on the set $L_2 ([0,\pi])$. More precisely, for
$L$ to be self-adjoint, we have to:
 
- either restrict it on functions satisfying
\be
\psi (0) = \lambda \psi (\pi)
\ee
where $\lambda$ plays the r\^{o}le of an anyonic parameter
($\lambda =1$ characterizes the bosons, while $\lambda = -1$
characterizes fermions)
 
- or modify the explicit form of $L$
\be
L=-i \frac{\prt}{\prt \phi} + \al
\ee
and apply it on functions such that
\be
\psi (0) = \psi (\pi)
\ee
the anyonic parameters being here $\al \ (\al =0$ being the bosons,
and $\al =1$ being the fermions).
 
\indent
 
Choosing the second alternative, which is compatible with the
bosonic case provided by the $x_i, p_j$ representation, leads to adding
a $\cw$ contribution to the $L$ defined in (\ref{eq:6.8}). Actually,
it appears possible to propose different $\cw$ realizations of the $sp(4,\R)$
algebra, i.e. different expressions including $\cw$ contributions to
the $sp(4,\R)$ generators; of course these different realizations
will be equivalent once a finite dimensional $\cw$
representation has been chosen. Among the different possibilities (see
ref. \cite{3}), the simplest one is the following:
\be
L' = -i \frac{\prt}{\prt \phi} -S
\ee
 
For the one-dimensional representations of the $\cw$ algebra, $S$
becomes a number that can be non-zero. For higher dimensional $\cw$
representations, $S$ is a diagonal matrix, and this framework could
lead to a generalization of anyons directly related to the enlarged
$\cw$ algebra (when compared with paragraph 6.1). A more complete
discussion of the validity of this interpretation can be found in
ref. \cite{3}.
 
As a conclusion, owing to a $\cw$ algebra treatment, the one-parameter 
self-adjoint extension family of the angular momentum does
allow the anyonic statistics to be incorporated.
 
\vs{0.5}
 
\sect{Conclusion and perspectives}
 
\indent
 
The characterization of the class of finite $\cw$ algebras that we
proposed, based on completely algebraic grounds, is conceptually
simple. Such an algebra is defined in terms of the commutant, in a
particular localization of the enveloping algebra $\cu(\cg)$, of a
subalgebra $\tilde{\cg}$ of a simple Lie algebra $\cg$. This approach
is specially adapted to obtain new realizations of the Lie algebra
$\cg$ from a particular (differential) one. In other words, a class
of $\cg$ representations can be explicitly built with the help of a
$\cg$ differential realization and the knowledge of a particular $W$
algebra. A direct comparison of this construction with the technique
of induced representations has been given in section 5 for the cases
of the four-dimensional conformal and Poincar\'e algebras. It was
also shown in section 6 that this framework fits with the Heisenberg
quantization for a system of two identical particles in two
dimensions, the $\cw$ algebra under consideration being
interpreted as carrying the anyonic information.
 
Therefore, it appears reasonable to put some more effort 
in the physical applications, as well as in the mathematical developments
of these objects. The first question that could be raised concerns the
Heisenberg quantization for a system of more than $N=2$ identical
particles. For $N \geq 3$, the structure of the algebra of
observables becomes much more complicated (see for example
ref.\cite{10}). The anyonic problem deserves more work, in
particular if we keep in mind the relevance of the $\cw_{1+\infty}$
algebra in the algebraic treatment of the quantum Hall effect (ref.
\cite{11}). As a second question, one could wonder about the
occurrence of a finite $\cw$ algebra as the symmetry algebra for a
particular Hamiltonian; several examples corresponding to different
types of potential have already been detected (see for example ref.
\cite{1}). It might be of some interest to bypass the mere
observation and try to understand how the non-linearity of the
symmetry algebra arises.
 
On the mathematical side, it looks promising to connect our approach
with the general study of primitive ideals considered in ref.
\cite{12}, in which the commutants of nilpotent algebras are
directly involved. But the most natural field of investigation is
of course the generalization of our constructions to the affine ---or
not finite--- $\cw$ algebra case.
 
Actually the realization of $\cw$ algebras with the generators
expressed in terms of all the generators of an affine Kac-Moody
algebra has been performed in ref. \cite{4}. As in the finite case,
the (primary) $\cw$ fields are written as quotients of
polynomials. However, we have to stress that the denominators, in these
quantities, 
 simply commute with all the numerators, allowing in particular
 a computation of the operator product expansions (OPE) without special
 difficulties  in  the  quantum  framework. Such a construction
 might be seen as a sort of generalized Sugawara one, but without the
restrictions to special values of the Kac Moody
central extension, and without need of coset technique, as developed
in ref. \cite{13}.
 
Finally, affine $\cw$ algebras could also be used to obtain realizations
of affine Kac-Moody
ones\cite{14}\cite{15}. Let us close this section by presenting such an
approach in the
$\widehat{sl}(2)_k$ case.
 
For such a purpose, we need first to define the $\widehat{sl}(2)_k$
currents $J_-(z), J_0(z)$ and $J_+(z)$ which satisfy the OPE's
\bea
J_0(z) J_\pm(w) &\sim& \pm \frac{1}{z-w} J_\pm (w) \; \; \; \; \;
\; \; \; \;  J_0(z) J_0(w) \sim \frac{k/2}{(z-w)^2} \nonumber \\
J_+(z) J_-(w) &\sim & \frac{k}{(z-w)^2} + 2 \frac{J_0 (w)}{z-w}
\ena
where the $\sim$ symbol indicates that we restrict the OPE to its
singular part.
 
Then, we introduce the $T(z)$ operator, which commutes with
$J_0(z)$ as well as $J_-(z)$:
\be
T=T_{sug} - \prt J_0 + : \frac{\prt J_-}{J_-} J_0 : + k \left[
\frac{3}{4} \left( \frac{\prt J_-}{J_-} \right)^2 - \frac{1}{2}
\frac{\prt^2J_-}{J_-} \right]
\ee
with $T_{sug} = \frac{1}{k+2} : J_0 J_0 + \shalf (J_+ J_- + J_-
J_+):$, and $: \; \; \; :$ being the normal ordered product.
 
$T(z)$ generates the Virasoro algebra with central charge
\be
c=1 -6\frac{(k+1)^2}{k+2}
\ee
 
In the following, we will use: $W(z) = (k+2) T(z)$, which for $k=-2$
reduces to: $J_0 J_0 + \shalf (J_+ J_- + J_- J_+):(z)$, and
commutes also with $J_+(z)$.
 
Now, as in the Wakimoto construction \cite{16}, we start with a
$(\beta, \gamma)$ system satisfying:
\be
\beta (z) \gamma(w) \sim \frac{1}{z-w}
\ee
which can
be considered as the ``affine'' analogue of the $(x, \prt_x)$ pair.
 The realization of $\widehat{sl}(2)_k$
we obtain reads:
\be
J_- = \gamma \; \; \; \; \; \; \; \; J_0 =-:\gamma \beta : -
\frac{k+2}{4} : \frac{\prt \gamma}{\gamma} :
\ee
\bea
J_+ =-: \gamma \beta^2 : - k \prt \beta &+& : \frac{1}{\gamma} W : +
(k+2) \left[ - \shalf :: \frac{\prt \gamma}{\gamma} : \beta :
\right. \nonumber \\
&+& \frac{1}{4} \left. \left( (k+1) \frac{\prt^2 \gamma}{\gamma^2} -
\frac{1}{4} (5k+6) \frac{(\prt \gamma)^2}{\gamma^3} \right) \right]
\ena
 
We could say that in our construction, the Wakimoto $\phi(z)$ free
field has been replaced by the Virasoro $W(z)$ operator. Fractional
calculus technique \cite{17} are well adapted to this framework,
which, we hope, might have its interest in the computation of correlation
functions.
 
By the Sugawara construction, we knew how to obtain, from an affine Lie
algebra, a Virasoro realization. The alternative to the Wakimoto
construction given just above allows us, starting from a Virasoro
representation, to obtain new $\widehat{sl}(2)_k$ ones: shall we
conclude that we have looped the loop?
 
\vs{0.5}
 
\noindent{\Large {\bf Acknowledgements}}
 
\indent
 
Paul Sorba is indebted to Prof. C. Burdik in Prague, Prof. M.
Eliashvili in Tbilisi, Prof. M.-L. Ge in Tianjin and their
collaborators for the opportunity given to him to present the
results and remarks gathered in this mini-review.
 
This work is partly supported by the European network contract No.
ERBCHRXCT920069.


\begin{thebibliography}{99}
\bibitem{1}J.~De~Boer, F.~Harmsze and T.~Tjin, Phys.
Report {\bf 272} (1996) 139.
 
\bibitem{2}L. Feher, L. O'Raifeartaigh, P. Ruelle, I. Tsutsui and
A. Wipf, Phys. Rep. {\bf 222} (1992) 1, and references therein.
 
\bibitem{3}F.~Barbarin, E.~Ragoucy and P.~Sorba, \NP{B442} (1995)
425.
 
\bibitem{4}F.~Barbarin, E.~Ragoucy and P.~Sorba, \IJMP{A11} (1996)
2835.
 
\bibitem{5}F.~Barbarin, E.~Ragoucy and P.~Sorba, {\em $\cw$
realization of Lie algebras: Application to $so(4,2)$ and Poincar\'e
algebras}, to appear in \CMP.
 
\bibitem{6bis}V.G.~Drinfel'd, Soviet Math. Dokl. {\bf 32} (1985)
254.
 
\bibitem{6}P.~Moussa and R.~Stora, in {\em Methods in Subnuclear
Physics}, Gordon and Breach, Herceg-Novi Summer School, 1966.
 
\bibitem{7}G.~Mack, \CMP{55} (1977) 1.
 
\bibitem{8} J.M. Leinaas and J. Myrheim, \IJMP{A8} (1993) 3649.
 
\bibitem{9} J.M. Leinaas and J. Myrheim, \PR{B37} (1988) 9286.
 
\bibitem{10} S.B.~Isakov and J.M. Leinaas, \NP{B463} (1996) 194.
 
\bibitem{11} A.~Cappelli, C.A.~Trugenberger and G.R.~Zemba, DFF
257/10/96; hep-th/9610019, {\em $W_{1+\infty}$ field theories for
the edge excitations in the quantum Hall effect}.
 
\bibitem{12} A.~Joseph, Journ. Algebra {\bf 48} (1977) 241.
 
\bibitem{13} F.A. Bais, P. Bouwknegt, M.~Surridge and K.~Schoutens,
\NP{B304} (1988) 348; ibid 371.
 
\bibitem{14} F.~Barbarin, {\em Alg\`ebres $W$ et applications},
th\`ese de doctorat, October 1996.
 
\bibitem{15} F.~Barbarin, E.~Ragoucy and P.~Sorba, in preparation.
 
\bibitem{16} M.~Wakimoto, \CMP{104} (1986) 605.
 
\bibitem{17} J.L.~Petersen, J.~Rasmussen and M.~Yu, NBI-HE-95-42,
hep-th/9512175, {\em Free field realization of $SL(2)$ correlators
for admissible representations, and Hamiltonian reduction for
correlators}.
 
\end{thebibliography}
\end{document}